\documentclass[aip,jmp]{revtex4-1}
\usepackage[dvips]{graphicx}
\usepackage{amsmath}
\usepackage{amssymb}
\usepackage{color}
\usepackage{bm}
\usepackage{url}

\newcommand{\commut}[2]{[#1,#2]}

\newcommand{\ket}[1]{|#1\rangle}

\newcommand{\brket}[3]{\langle#1|#2|#3\rangle}

\newcommand{\overlap}[2]{\langle#1|#2\rangle}
\newcommand{\avg}[1]{\langle#1\rangle}
\newcommand{\tm}[1]{\textrm{#1}}
\newcommand{\kk}[0]{\bm{k}}

\begin{document}


\title{Approach to Equilibrium of a Restricted Class of Isolated Quantum Systems After a Quench}


\author{E. Solano-Carrillo}
\affiliation{Department of Physics, Columbia University, New York, NY 10027, USA}


\begin{abstract}
We prove the approach to equilibrium of quenched isolated quantum systems for which the change in the Hamiltonian brought about by the quench satisfies a certain closed commutator algebra with all the extensive integrals of motion of the system before the quench. The proof is carried out by following the exact unitary evolution of the entropy operator, defined as the negative of the logarithm of the nonequilibrium density matrix, and showing that, under the conditions implied by the assumed algebra, this entropy operator becomes, at infinite times, a linear combination of integrals of motion of the perturbed system. That is, we show how the nonequilibrium density matrix approaches a generalized Gibbs ensemble. The restricted class of systems for which the present results apply turn out to have degenerate spectra in general, as opposed to some generic systems for which a kind of ergodicity is expected, with a nontrivial dynamics believed to find instances in one-dimensional infinite super-integrable systems. Our findings constitute a direct demonstration of how a non-ergodic isolated quantum system may get to statistical equilibrium. 
\end{abstract}


\maketitle

\section{Introduction}
It is of great interest at present to understand what happens, after a sufficiently long time, when an isolated quantum many-body system which is prepared in a certain initial state is suddenly altered (quenched). This situation can be studied experimentally, e.g. in ultra-cold atomic gases \cite{Kinoshita,Hofferberth,Trotzky,Gring,Cheneau} where a high degree of isolation can be achieved as well as an unprecedented tunability of the interactions. The experiments show that some observables approach stationary values that are consistent with the expectations taken with equilibrium statistical ensembles. This has triggered a considerable amount of theoretical research \cite{Polkovnikov,Yukalov,Eisert,Calabrese}, and it is a common belief now that generic systems thermalize \cite{Deutsch,Srednicki,Rigol2} in the sense that they approach to equilibrium regardless of the initial states, while integrable systems attain stationary values described by generalized Gibbs ensembles \cite{Rigol}, where the system remembers the initial conditions by means of all nontrivial integrals of motion.

Following the spirit of statistical mechanics, a substantial theoretical activity is switching toward understanding the most appropriate statistical ensembles describing the long-time behavior of each system of interest instead of solving the complicated nonequilibrium dynamics. This then takes for granted that the asymptotic behavior of the system can well be described by a steady-state quantum statistical ensemble so if this representation is to be \emph{unique} and an ensemble description is to be unambiguously valid at all times, as originally advocated by Gibbs \cite{Gibbs}, the natural question arises of how a trajectory in the space of probability density operators can lead to a steady-state operator for isolated systems, defined as those whose density matrices evolve according to the Liouville-von Neumann equation.

In order to fully justify the current trend to understand the nonequilibrium dynamics of isolated quantum systems, it therefore remains to be shown, and this was briefly discussed by the author in \cite{Solano1}, whether it is possible that the unitary evolution of an isolated quantum system can bring the density matrix along a path whose asymptotics coincides with a steady-state density matrix. Attempts to show this on a general basis \cite{Hershfield,Dutt} have required the assumption of the existence of a physical relaxation process in the system causing correlation functions to decay in time. Although this might be the case for special closed systems for which some parts serve as effective reservoirs for the other parts, which is not always the case as exemplified by systems with many-body localization \cite{Nandkishore}, an exact derivation free of these kind of assumptions is of considerable interest.

Since for a finite system the density matrix rotates as a vector in Liouville space giving rise to quantum recurrences \cite{Solano1}, it is clear that such a possibility of having a steady-state quantum statistical ensemble as a result of a unitary dynamics may arise only for infinite systems. This is substantiated by the known fact that the sole operation of taking the thermodynamic limit in an isolated many-body systems may lead \cite{MehtaA} to an effective dissipation mechanism (quasiparticles dissipating energy infinitely far away), making, e.g., the poles in the self-energy to merge into a branch cut. 

Remarkably, it is precisely the limit of infinitely large systems and the presence of self-energy effects that has allowed in the past the formulation of a rigorous theory for the approach to equilibrium in quantum statistics \cite{vHove,PrigRes,Res,Fujita,Nakajima,Zwanzig2}, where off-diagonal oscillating terms in the density matrix are shown to be negligibly small \cite{PrigHen,Henin} after evaluating the asymptotic time integrals involved when a diagramatic expansion with respect to the perturbation deviating the system from equilibrium is performed, instead of just a representation of the density matrix in the basis of eigenstates of the perturbed Hamiltonian.

In this paper we identify sufficient conditions for an \emph{exact} proof of the approach to equilibrium in a certain kind of infinite isolated quantum systems after a quench. These are prepared in initial equilibrium states of the generalized Gibbs exponential form and are subject to a sudden change in their Hamiltonian which satisfies a special algebraic property (stated in \eqref{pV}) that substantially restricts the range of applications to physical systems usually found in practice but nonetheless has the effect, at least from a strict mathematical sense, of making the density matrices satisfying the Liouville-von Neumann equation asymptotically approach different equilibrium states that are also of the generalized Gibbs exponential form. The proof uses the unitary evolution of the entropy operator \cite{Solano2}, defined as the negative of the logarithm of the density matrix, from which a theory of entropy production in quantum many-body systems can be established as a quantum-mechanical basis for nonequilibrium thermodynamics. 

We organize the discussion as follows: in section \ref{s1} we introduce the initial states and the property, \eqref{pV}, of the quench from which the equilibration of the isolated quantum systems follows; then we proceed in section \ref{s2} to the proof of the approach of the entropy operator to a linear combination of integrals of motion of the quenched system at infinite times. A discussion is made in section \ref{s3} of the consequences of \eqref{pV} regarding the general nature of the systems and quenches for which the present results apply, specifically the degeneracy of the unperturbed stationary states, and a comparison with the equilibration on average of some generic quantum system is made. We conclude with section \ref{s4}.

\section{Initial states and quench protocol}\label{s1}
We consider an isolated system with Hamiltonian $\hat{H}_0$, which at time $t=0$ is in statistical equilibrium, described by the density matrix
\begin{equation}\label{r0}
 \hat{\rho}_0=\exp\left(-\textstyle\sum_n\hat{S}_n^0/T\right),
\end{equation}
where the $\hat{S}_n^0$'s are an independent set of integrals of motion of the dynamics generated by $\hat{H}_0$, i.e. 
\begin{equation}\label{c0}
 \commut{\hat{H}_0}{\hat{S}_n^0}=0,
\end{equation}
and $T$ is the temperature of a reservoir in contact with the system for times $t\le0$. The type of initial states embodied by \eqref{r0} include, e.g., the Gibbs grand canonical ensemble, in which case 
\begin{equation}\label{Sng}
\hat{S}_1^0=\hat{H}_0, \hspace{0.5cm} \hat{S}_2^0=-\mu \hat{N}, \hspace{0.5cm}\hat{S}_3^0=-\Omega\, \hat{1},
\end{equation}
with $\mu$ being the chemical potential of the group of particles comprising the system, with number operator $\hat{N}$, and $\Omega$ the grand potential.
 
We shall use in this paper the term generalized Gibbs ensemble (GGE) when the set of integrals of motion has, in general, more elements than described above, with the Gibbs ensemble being just a particular case. Having fixed this number of integrals of motion, $D$, we call
\begin{equation}\label{S0t}
 \hat{S}_0=\dfrac{1}{T}\sum_{n}\hat{S}_{n}^0=\dfrac{1}{T}\,\bm{\theta}^0\cdot \hat{\bm{S}}^0,
\end{equation}
the initial \emph{entropy operator}, with $\bm{\theta}^0$ the $D$-dimensional vector of ones, and $\hat{\bm{S}}^0$ the $D$-dimensional vector with the $n^{\textrm{th}}$ component given by $\hat{S}_{n}^0$. We follow the convention from hereon that $\hat{S}_1^0=\hat{H}_0$ and $\hat{S}_D^0=-\Omega\,\hat{1}$ with $\exp(-\Omega/T)=\tm{Tr}\,\exp(-\sum_{n\neq D}\hat{S}_n^0/T)$. Our initial equilibrium states, represented by density matrices of the form
\begin{equation}\label{r0s}
  \hat{\rho}_0=\exp(-\hat{S}_0),
\end{equation}
are then a subset of those mixed states which can be obtained as a result of equilibration after the system was manipulated in the remote past for the sake of initial preparation.

At time $t=0$ the reservoir is removed, while the system is suddenly altered by a perturbation $\hat{V}$ and is left alone evolving with the new Hamiltonian $\hat{H}=\hat{H}_0+\hat{V}$. The defining property of the restricted class of systems and quenches dealt with here is that the following commutator algebra, which we term closed, is satisfied
\begin{equation}\label{pV}
 i\commut{\hat{V}}{\hat{S}_n^0}=\sum_{m}\Gamma_{nm}\,\hat{S}_m^0,
\end{equation}
with $\Gamma$ a symmetric and positive semi-definite $D\times D$ real matrix with finite norm and having the eigenvalue zero with a degeneracy greater than one. We now want to derive from \eqref{pV} some consequences regarding the nature of the systems and the quenches.

Evaluating \eqref{pV} at $n=1$ and using the Heisenberg equation of motion for the perturbing potential $\hat{V}$ at time $t=0$, that is, $(d\,\hat{V}_H/dt-\partial\hat{V}_S/\partial t\,)_{t=0}=i\commut{\hat{H}_0}{\hat{V}_S}\mid_{t=0}$, where the subscripts $H$ and $S$ denote the Heisenberg and Schr\"{o}dinger pictures, respectively, we have
\begin{equation}\label{delta}
 2 (d\hat{V}_H/dt)_{t=0}-2\hat{V}\,\tm{lim}_{t\rightarrow0}\,\delta(t)=-\sum_m \Gamma_{1m}\,\hat{S}_m^0.
\end{equation}
To obtain this we have written $\hat{V}_S(t)=\hat{V}\Theta(t)$ due to the singular nature of the quench, where $\Theta(t)$ is the Heaviside step function, defined such that $\Theta(0)=1/2$.

The left hand side of \eqref{delta} does not vanish identically provided $\commut{\hat{H}_0}{\hat{V}}\neq0$. This is actually the case in the present discussion since we discard the trivial case of the perturbing potential being an integral of motioin itself, which therefore implies (adding quantities of the same order of magnitude) that 
\begin{equation}
(d\hat{V}_H/dt)_{t=0}\equiv \hat{Q}\,\tm{lim}_{t\rightarrow0}\,\delta(t), 
\end{equation}
where $\hat{Q}$ is an operator representing the \emph{total} energy interchanged between the system and the reservoir in the operation of turning on the perturbing potential, with $\commut{\hat{H}_0}{\hat{Q}}\neq0$. Since this is arbitrarily determined in part by the nature of the reservoir in contact with the system for times $t\le0$ as well as by some features of the system itself (e.g. its heat capacity), we can assume, without loss of generality, that 
\begin{equation}\label{unc}
 i\commut{\hat{Q}}{\hat{S}_n^0}=\delta_{n1}\,\sum_{m}\Gamma_{1m}\hat{S}_m^0,
\end{equation}
From the point of view of measurement theory, \eqref{unc} just establishes an uncertainty principle expressing the incompatibility of simultaneously measuring the initial internal energy of the system \emph{and} the total energy interchanged with the reservoir at $t=0$ by turning on the perturbing potential. Simultaneous measurement of the latter with each integral of motion is however possible according to \eqref{unc}. Introducing the operator representing the speed at which $\hat{Q}$ is interchanged 
\begin{equation}\label{Qd}
\dot{\hat{Q}}= i\commut{\hat{H}_0}{\hat{Q}}=-\sum_m\Gamma_{1m}\hat{S}_m^0,
\end{equation}
we see that, unless $\hat{Q}$, measuring $\dot{\hat{Q}}$ is compatible with the measurement of the internal energy of the system before the quench, which is the reason why \eqref{unc} is assumed in the first place, since in this way the thermodynamics of the quenching procedure is well defined.

Three important conditions are then \emph{derived} from \eqref{delta} regarding the nature of our restricted class of systems and quenches:
\begin{itemize}
 \item The systems must be infinite and one-dimensional. 
 \item The perturbing potential must be local.
 \item The integrals of motion must form a Lie algebra.
\end{itemize}
That the systems must be infinite and one-dimensional can be deduced from \eqref{delta} if we remind that equilibrium statistical states are described in terms of \emph{extensive} integrals of motion, i.e. those proportional to the volume $\nu_d$ of the system, with $d$ denoting the dimensionality. When a mode expansion ($\sum_{\kk} \cdots$) is made for operators and the volume of the system is sent to infinity, it is known that Kronecker deltas become Dirac deltas \cite{Mahan} according to  $\delta_{\kk,\kk'}=\tm{lim}_{\nu_d\rightarrow\infty}((2\pi)^{d}/\nu_d) \,\delta^{(d)}(\kk-\kk')$, from which it is seen that $\tm{lim}_{\nu_d\rightarrow\infty}\,\nu_d \propto \tm{lim}_{\kk\rightarrow \kk'}\delta^{(d)}(\kk-\kk')$. Therefore, in order for \eqref{delta} to make sense, the integrals of motion must be unbounded as a result of taking the limit of infinite systems. Moreover, for the singularities in both sides of \eqref{delta} to be of the same order, it is required that $d=1$, i.e. one-dimensional systems.

Denoting with $L$ the length of the system, the integrals of motion must then be proportional to $\tm{lim}_{L\rightarrow\infty}L=2\pi\, \tm{lim}_{k\rightarrow k'}\,\delta(k-k')=(2\pi a^2/v_c)\,\tm{lim}_{t\rightarrow0}\,\delta(t)$ where, by mere dimensional analysis, $a$ is a typical microscopic length scale (e.g. typical interparticle distance) and $v_c$ a typical velocity (e.g. velocity of propagation of correlations). In this way  the singularities in \eqref{delta} cancell in both sides and we can express the perturbation $\hat{V}$ in terms of the integral densities as
\begin{equation}\label{defV}
 \hat{V}=\hat{Q}+\dfrac{\pi a^2}{v_c}\sum_m \Gamma_{1m}\,\hat{s}_m^0,\hspace{0.5cm}\hat{s}_m^0\equiv \hat{S}_m^0/L.
\end{equation}
Since the initial and final energies of the system depend on its size whereas the change is size-independent, we conclude that the total energy added to (or removed from) the system due to the quench is relatively very small and hence the perturbation must be local. 

Now, that the integrals of motion must form a Lie algebra can easily be inferred after substituting \eqref{defV} into \eqref{pV}. By doing this, we obtain
\begin{equation}\label{cm}
 i\commut{\hat{Q}}{\hat{S}_n^0}+ \dfrac{\pi a^2}{v_c}\sum_m \Gamma_{1m}\,i[\hat{s}_m^0,\hat{S}_n^0]=\sum_{l}\Gamma_{nl}\hat{S}_l^0.
\end{equation}
Using \eqref{unc} we see that for this expression to hold it is sufficient and necessary that the following commutation relations are satisfied
\begin{equation}\label{lie}
 [\hat{S}_n^0,\hat{s}_m^0]=i\sum_{l}\Lambda_{nm}^l \hat{S}_l^0,
\end{equation}
with $\Lambda_{mn}^l$ the so-called structure constants of the Lie algebra, with $\Lambda_{nm}^l=-\Lambda_{mn}^l$ and, by definition, $\Lambda_{1m}^l=0$. Substituting this in \eqref{cm} and using the linear independence of the integrals of motion we get the relations
\begin{equation}\label{defG}
 \Gamma_{nl}= \dfrac{\pi a^2}{v_c}\sum_m\Gamma_{1m}\Lambda_{nm}^l,\hspace{0.5cm} n\neq1.
\end{equation}
Note that in order to have a symmetric $\Gamma$ matrix, it is required that $\Lambda_{nm}^l=\Lambda_{lm}^n$. Eq. \eqref{defG} completes the formulation of the kind of systems and quenches of interest here in terms of the structure constants of the Lie algebra satisfied by the extensive integrals of motion and the parameters $\Gamma_{1m}$, which are to be taken in such a way that the matrix $\Gamma$ is positive semi-definite and has a degenerate eigenvalue zero. The physical interpretation of these parameters is fixed by \eqref{Qd}, in terms of which we can rewrite \eqref{defV} as
\begin{equation}
 \hat{V}=\hat{Q}-\dfrac{\pi a^2}{v_c L}\dot{\hat{Q}}.
\end{equation}
That is, of the total energy interchanged between the system and the reservoir at $t=0$, a part represented by $(\pi a^2/v_c L)\,\dot{\hat{Q}}$ goes to the reservoir, and the remaining, represented by $\hat{V}$, is in charge of the change in the internal energy of the system. Since simultaneous measurement of $\hat{Q}$ and $\dot{\hat{Q}}$ are incompatible, as expressed by the commutation relation $i\commut{\hat{Q}}{\dot{\hat{Q}}}=\Gamma_{11}\dot{\hat{Q}}$, the more information we have about the total energy exchanged, the least information we know about that absorbed by the reservoir so $\dot{\hat{Q}}$ may be interpreted as an operator representing a kind of unavoidable waste heat resulting from turning on the perturbing potential.

Regardless of whether or not a system and a quench are easily found in practice to obey the above properties, our aim is to show that the subsequent unitary evolution leads, without invoking a dissipation mechanism, to equilibration. Furthermore, we show that the equilibrium state attained after the quench is a GGE with the integrals of motion of the dynamics generated by the final Hamiltonian $\hat{H}$.

\section{Proof of the approach to equilibrium}\label{s2}
The density matrix for times $t\ge0$ can be conveniently written as \cite{Solano2}
\begin{equation}\label{rt}
 \hat{\rho}_t=\exp(-\hat{S}_t),
\end{equation}
where the entropy operator $\hat{S}_t$ satisfies, just as the density matrix does, the von Neumann equation, whose solution is 
\begin{equation}\label{eet}
 \hat{S}_t=e^{-i\hat{H}t}\hat{S}_0e^{i\hat{H}t}.
\end{equation}
This can be easily verified by substituting \eqref{eet} in \eqref{rt}, expanding in a power series, and realizing that we get the known unitary evolution $\hat{\rho}_t=e^{-iHt}\hat{\rho}_0e^{iHt}$. The expectation value of the entropy operator gives the von Neumann entropy, however, it is only the thermodynamic entropy operator, obtained from $\hat{S}_t$ by taking its diagonal part in the energy representation, which can be related to a nonequilibrium entropy in accordance with thermodynamics \cite{Solano2}. 

To get a simple derivation of our results, we introduce the notation $\mathcal{L}_{A}\equiv \commut{\hat{A}}{\cdot}$ for Liouvillians. We can then write \eqref{c0} and \eqref{pV}, in matrix notation, as
\begin{equation}\label{lV}
\begin{split}
 \mathcal{L}_{{H_0}}\hat{\bm{S}}^0&=0,\\
i\mathcal{L}_V\hat{\bm{S}}^0&=\Gamma\cdot\hat{\bm{S}}^0.
\end{split}
\end{equation}
Also, \eqref{eet} can be written in Liouvillian notation as 
\begin{equation}
 \hat{S}_t = e^{-i\mathcal{L}_Ht}\hat{S}_0.
\end{equation}
The proof now follows by using the Dyson decomposition, valid also for Liouvillians \cite{Zwanzig},
\begin{multline}
  e^{-i\mathcal{L}_Ht}=1+(-i)\int_0^tdt_1\,e^{i\mathcal{L}_{H_0}(t_1-t)}\mathcal{L}_Ve^{-i\mathcal{L}_{H_0}t_1}\\+(-i)^2\int_0^tdt_1\int_0^{t_1}dt_2 \,e^{i\mathcal{L}_{H_0}(t_1-t)}\mathcal{L}_Ve^{i\mathcal{L}_{H_0}(t_2-t_1)}\\
\times\mathcal{L}_Ve^{-i\mathcal{L}_{H_0}t_2}+\cdots.
\end{multline}
By using \eqref{lV} and $\int_0^tdt_1\int_0^{t_1}dt_2\cdots\int_0^{t_{k-1}}dt_k=t^k/k!$, we obtain for the entropy operator
\begin{equation}
 \hat{S}_t=\dfrac{1}{T}\,\bm{\theta}^0\cdot e^{-\Gamma t}\cdot \hat{\bm{S}}^0.
\end{equation}
Since $\Gamma$ is real and symmetric, it is diagonalized by an orthogonal transformation $M$,
\begin{equation}\label{ei}
 M^T\cdot \Gamma \cdot M=\textrm{diag}(\gamma_1,\cdots,\gamma_D),
\end{equation}
where $T$ denotes transpose. Introducing the transformed vectors
\begin{equation}\label{tv}
 \bm{\theta}^e=M^T\cdot \bm{\theta}^0,\hspace{0.3cm}\textrm{and}\hspace{0.3cm}\hat{\bm{S}}^{e}=M^T\cdot\hat{\bm{S}}^0,
\end{equation}
we can rewrite the entropy operator as
\begin{equation}\label{et}
  \hat{S}_t=\dfrac{1}{T}\sum_n\theta_n^e\,e^{-\gamma_n t}\,\hat{S}_n^{e},
\end{equation}
where the $\gamma_n$'s are the (non-negative) eigenvalues of $\Gamma$. 

\subsection{GGE of the quenched system}
If we label the subset of zero eigenvalues of $\Gamma$ with the index $r$, we have from \eqref{rt} and \eqref{et} in the infinite-time limit
\begin{equation}
 \hat{\rho}_{\infty}=\exp\left(-\textstyle\sum_r\theta_r^e\hat{S}_r^e/T\right).
\end{equation}
We now prove that the $\hat{S}_r^e$'s are integrals of motion of the quenched system, i.e. 
\begin{equation}\label{cHSe}
\commut{\hat{H}}{\hat{S}_r^e}=0. 
\end{equation}
From \eqref{lV} and \eqref{tv} we have 
\begin{equation}\label{lH}
 i\mathcal{L}_H \hat{S}_r^e = i\mathcal{L}_V(M^T\cdot\hat{\bm{S}}^0)_r=\sum_n(M^T\cdot\Gamma)_{rn}\,\hat{S}_n^0.
\end{equation}
Using \eqref{ei} in the form $M^T\cdot\Gamma= \textrm{diag}(\gamma_1,\cdots,\gamma_D)\cdot M^T$ we get
\begin{equation}
(M^T\cdot\Gamma)_{rn}=\gamma_rM_{nr}=0,
\end{equation}
the last equality because $\gamma_r=0$ by definition. Substituting in \eqref{lH} we get the desired result, \eqref{cHSe}. This shows that the steady-state density matrix (i.e. $\commut{\hat{H}}{\hat{\rho}_{\infty}}=0$) is a GGE as defined here. The number of integrals of motion of the quenched system, which is the same as the number of zero eigenvalues of $\Gamma$ is then, in the nontrivial cases where $\commut{\hat{V}}{\hat{H}_0}\neq0$, less than in the unperturbed system, which expresses the fact that $\hat{V}$ has to be necessarily a symmetry-breaking perturbation.

\section{Discussion}\label{s3}
We stress that the conclusions derived here required the systems to be infinite, with the integrals of motion being unbounded operators as a result.  Failure to comply with this lead immediately to contradictory results as can be seen if we multiply \eqref{pV} by $M_{nl}$ and sum over $n$, so that we can rewrite the assumed commutator algebra, \eqref{pV}, in the simplified form
\begin{equation}\label{ol}
 i[\hat{V},\hat{O}_l]=\gamma_l \hat{O}_l, 
\end{equation}
where the operators $\hat{O}_l=\sum_{n=1}^DM_{nl}\hat{S}_n^0$ are hermitian and different from the null operator. Now, our results are nontrivial as long as $\Gamma$ has at least one nonzero eigenvalue, say $\gamma_j>0$. Taking the expectation value of \eqref{ol}, with $l=j$, in an eigenstate of $\hat{O}_j$ with nonzero eigenvalue
\begin{equation}\label{op}
\hat{O}_j \ket{\Psi}=\lambda \ket{\Psi},\hspace{1cm} \lambda\neq0,
\end{equation}
the following is then obtained
\begin{equation}\label{ab}
 0= i\brket{\Psi}{[\hat{V},\hat{O}_j]}{\Psi}=\gamma_j \brket{\Psi}{\hat{O}_j}{\Psi}=\gamma_j \lambda \overlap{\Psi}{\Psi}.
\end{equation}
To understand the pathology of this equation, we note that it is of the same nature of the result obtained when the expectation value of the canonical commutation relations for the position and momentum operators, $\hat{q}$ and $\hat{p}$ respectively, of a particle with one degree of freedom
\begin{equation}
 [\hat{q},\hat{p}]=i\hbar\,\hat{1},
\end{equation}
is taken in an eigenstate, say, of the position operator $\hat{q}$, with non-zero eigenvalue
\begin{equation}\label{ab2}
 0=\brket{q}{[\hat{q},\hat{p}]}{q}=i\hbar\,\overlap{q}{q}.
\end{equation}
The contradiction does not imply, of course, that the canonical commutation relations are invalid. They just do not have solution for finite-dimensional spaces (unless $\hbar=0$), or for either $\hat{q}$ or $\hat{p}$ being \emph{bounded} operators \cite{Weyl,Rosenberg}. This is the same situation that we encounter here and, as with the well-known case of the states $\ket{q}$ not being normalizable and then invalidating \eqref{ab2}, the unbounded property of the operators $\hat{O}_l$ means that the vectors $\hat{O}_l\ket{\Psi}$ in \eqref{op} is not normalizable, making \eqref{ab} not well-defined. 

It is also important to remark that, in order to have a nontrivial dynamics, the integrals of motion of the unperturbed system should not be in involution or mutually commuting (as they are e.g. in integrable systems). This is the essence of the Lie algebraic relations \eqref{lie}, and points toward systems with \emph{degenerate} spectra such as those with accidental degeneracies or hidden symmetries, which are examples of a more general class of systems called super-integrable \cite{Miller}, which allow more integrals of motion than degrees of freedom. The study of quench dynamics in nonabelian integrable models has been pioneered by Fagotti and Bertini \cite{Fagotti,Bertini}.

We contrast the present derivation of the approach to equilibrium with the equilibration \emph{on average} suggested originally by von Neumann \cite{vNeumann} for quantum systems with \emph{nondegenerate} spectra (generic or typical systems), which may exhibit a kind of ergodic behavior when the time average of the expectation value of a few-body observable 
\begin{equation}
 \overline{\avg{\hat{A}(t)}}=\lim_{\tau\rightarrow\infty}\dfrac{1}{\tau}\int_0^\tau\avg{\hat{A}(t)}dt,
\end{equation}
coincides with its expectation value taken with a microcanonical ensemble. By expressing the nonequilibrium density matrix in the basis, $\lbrace \ket{i}\rbrace$, of stationary states of the perturbed system, satisfying $\hat{H}\ket{i}=E_i\ket{i}$,
\begin{equation}
 \overline{\avg{\hat{A}(t)}}=\sum_{ij}\sigma_{ij}\brket{j}{\hat{A}}{i}\Delta(\omega_{ij}),
\end{equation}
where $\omega_{ij}=E_i-E_j$ and the $\sigma_{ij}$'s are the matrix elements of the initial density matrix. The condition of the perturbed system having nondegenerate spectra, meaning $\omega_{ij}=0$ if and only if $i=j$, implies that
\begin{equation}
 \Delta(\omega_{ij})\equiv \lim_{\tau\rightarrow\infty}\dfrac{1}{\tau}\int_0^\tau \exp(-i\omega_{ij}t)dt=\delta_{ij},
\end{equation}
and then the time average becomes the expectation value taken with a diagonal ensemble
\begin{equation}\label{de}
 \overline{\avg{\hat{A}(t)}}=\sum_{i}\sigma_{ii}\brket{i}{\hat{A}}{i}.
\end{equation}
If the matrix elements $A_{ii}=\brket{i}{\hat{A}}{i}$ change slowly with the state $\ket{i}$, with $A_{i+1,i+1}-A_{ii}$ as well as $\brket{j}{\hat{A}}{i}$ ($i\neq j$) being exponentially small in the number of particles in the system, the assumption being known as the eigenstate thermalization hypothesis \cite{Deutsch,Srednicki,Rigol2,Rigol3}, then \eqref{de} can be argued to coincide with the expectation value taken with a microcanonical ensemble.

We see that the equilibration of the restricted class of infinite systems presented here is much stronger than that for generic systems (equilibration on average), since it represents the approach to a steady state of the entire density matrix. This approach contemplates the process of equilibration, in Gibbs' spirit, as a property of the whole statistical ensemble used to describe the system at all times and makes it strongly dependent on how the initial conditions were prepared.

\section{Conclusion}\label{s4}
We have shown that there exists, at least from a strictly mathematical viewpoint, a class of infinite isolated quantum systems prepared in equilibrium states for which the path followed by the nonequilibrium density matrix, after the system is quenched, can be proved to lead, at infinite times, to equilibrium generalized Gibbs ensembles. For this, it is sufficient to assume (and this defines the class) that the change in the Hamiltonian brought about by the quench satisfies a certain closed commutator algebra with all the extensive integrals of motion of the unperturbed system. Nontrivial cases are expected to occur in one-dimensional infinite systems exhibiting super-integrability. 

\section*{Acknowledgments}
\vspace*{0.5cm}
I am grateful for the support from the Fulbright-Colciencias Fellowship and the Columbia GSAS Faculty Fellowship.
\bibliographystyle{apsrev}
\bibliography{references}

\end{document}